\documentclass[a4paper]{article}
\usepackage[dvips]{graphicx}

\begin{document}

\title {\bf AC loss in ReBCO pancake coils and stacks of them: modelling and measurement}
\markboth{AC loss in stacks of pancake coils ...}{E. Pardo {\it et al.}}

\author{E. Pardo$^*$, J. {\v Souc}, J. {Kov\'a\v c}\\Institute of Electrical Engineering, Slovak Academy of Sciences,
\\Dubravska 9, 84104 Bratislava, Slovakia\\
$^*$ enric.pardo@savba.sk}

\date{}

\maketitle

\begin{abstract}
Many applications of $Re$BCO coated conductors contain stacks of pancake coils. In order to reduce their high AC loss, it is necessary to understand the loss mechanisms. In this article, we measure and simulate the AC loss and the critical current, $I_c$, in stacks of pancake coils (``pancakes"). We construct stacks of up to 4 pancakes and we measure them by electrical means. We also obtain the anisotropic field dependence of $J_c$ from $I_c$ measurements of the tape. This $J_c$ is the only input to the simulations, together with the coil dimensions. After validating our computations with the measurements, we simulate stacks of many pancakes, up to 32. We found that the AC loss in a stack of (four) pancakes is very high, two orders of magnitude larger than for a single tape. A double pancake behaves as a single one with double width but a stack of more pancakes is very different. Finally, we found that a 2-strand Roebel cable reduces the AC loss in a stack of pancakes but not in a single pancake. In conclusion, our simulations are useful to predict the ac loss of stacks of coated conductor pancake coils and to reduce the ac loss by optimizing the coil design.
\end{abstract}













\section{Introduction}

$Re$Ba$_2$Cu$_3$O$_{7-x}$ (or $Re$BCO) coated conductors (where $Re$ is a rare earth, usually Y, Gd or Sm) are the most promising high-temperature superconductors for applications because of two reasons: their high operation temperature, up to liquid nitrogen, and their good in-field performance. Many power applications contain windings, consisting on either solenoids or stacks of pancake coils. The latter is usually more convenient for winding technology because of the tape geometry of the coated conductors. Applications containing these windings are either AC (transformers, motors and generators, including wind generators) or DC (high-field magnets, superconducting magnetic storage systems). The AC loss in present AC devices is too high, specially for transformers \cite{iwakuma01IES,iwakuma09PhC} and the stator of rotating machines. Although using DC superconducting coils in the rotor already provides many advantages for superconducting rotating machines \cite{gieras09PET}, superconducting stators with sufficiently low AC loss further increase the efficiency and power per unit weight in all-superconducting motors \cite{kajikawa09IES}. In addition, using superconducting windings in the stator only, with permanent magnets in the rotor, requires a much simpler cryogenic system \cite{liL11IES,oswald11eucas}. Actually, the DC winding of the rotor is under the influence of a certain ripple magnetic field from the stator; that is, a small AC field on a DC background. This may cause significant AC loss, which should be taken into account in the cryogenic design. Reduction of AC loss in DC devices is also desirable because it limits the ramp speed. Moreover, hysteresis loss is often not negligible, specially for liquid-helium cooled systems. In order to reduce the AC loss, it is necessary to understand the loss mechanisms. In addition, measurements are required to characterise the windings and simulation tools are needed to predict the AC loss and design the cryogenic system. However, there is very little work on stacks of $Re$BCO pancake coils and, particularly, no work at all on AC loss simulations. Actually, detailed AC loss simulations are also missing for stacks of pancakes made of Bi$_2$Sr$_2$Ca$_2$Cu$_3$O$_{10}$ (Bi2223) tapes (the previous generation of high-temperature superconducting tapes), in contrast to the extensive experimental work \cite{bigoni00MPB,polak01IES,oomen03SST,tanaka04PhC}. In addition, there are no published simulations using a realistic anisotropic field dependence for the critical current density, $J_c$.

There are practically no AC loss measurements for stacks of pancake coils. Most of the work is for single pancake coils, both for coated conductors with ferromagnetic substrate \cite{polak06APL,pancsubs,pancakeFM} and non-magnetic one \cite{grilli07SST,kimJH11IESb}, or double pancakes with ferromagnetic substrate \cite{polak08IES,ainslie11SST}. The only measurements for a stack of pancake coils is \cite{rey11IESb}, where they use coated conductor with ferromagnetic substrate. The mechanisms of the AC loss in that work are not clear, presumably dominated by the ferromagnetic or eddy current contributions at low AC amplitudes. Other articles on stacks of pancakes measure only the critical current \cite{hase07IES,rey07IES}.

There are no AC loss simulations on stacks of pancake coils beyond a double pancake. There is, however, an extensive work on pancake coils made of coated conductor with both non-magnetic substrate \cite{grilli07SST,claassen06APL,clem07SSTb,pancaketheo,yuanW09SST,prigozhin11SST} or magnetic one \cite{ainslie11SST}. In addition, \cite{roebelcomp,ainslie11SST} calculate the AC loss in double pancakes (\cite{roebelcomp} discusses a Roebel cable but all the results are identical to those for a double pancake with the same cross-sectional structure as the cable). The only work on stacks of pancake coils predicts the critical current and the magnetic field but not the AC loss \cite{limS05SST}. The most relevant works comparing the simulations with experiments are \cite{yuanW11IES,simHTS2010}, about single pancakes.

Circular pancake coils and stacks of them are usually applied to transformers and high-field magnets. Concerning motors and generators, most prototypes contain racetrack coils \cite{gieras09PET}. However, the AC loss in circular coils and parallel stacks of strips are of the same order of magnitude, presenting the same qualitative features \cite{yuanW10SST}. Since racetrack coils contain circular and straight parts, they present the same qualitative behaviour as circular coils. Then, the results for circular coils are also useful to understand the loss mechanisms in racetrack coils and interpret the measurements.

In this article, we measure (in liquid nitrogen bath) and numerically simulate the AC loss and critical current in pancake coils and stacks of them made of coated conductor with non-magnetic substrate. For all cases, the coils are in self-field; that is, there is no magnetic field created by external sources. First, we measure stacks consisting on 1, 2, 3 and 4 pancakes practically identical to each other. Afterwards, we simulate these stacks and check the validity of the simulations by comparing to the measurements. The only input to the simulations are the coil dimensions and the anisotropic field dependence of the critical current density, $J_c$. We obtain this dependence from critical-current measurements of a tape sample. Finally, we simulate situations difficult to construct. These are the effects of stacking many pancakes, up to 32, and the reduction of the AC loss by replacing the single tape by a 2-strand Roebel cable.

This article is structured as follows. Section \ref{s.meas} details the experimental method and presents the measured critical current and AC loss of the 4 individual pancakes and stacks made of them. That section also presents the anisotropic field dependence of the critical current of the tape. Section \ref{s.sim} describes the simulation technique and discusses the results for the magnetic field, current distribution, critical current and AC loss, discerning the contribution from each pancake of the stack. In that section, we further discuss the experiments in light of the simulations of the experiments and other configurations which are difficult to measure. Finally, section \ref{s.concl} summarises our conclusions.


\section{Measurements}
\label{s.meas}

In the following, we present the measurements. First, we outline the experimental method, consisting entirely on electrical means, and the technique to construct the stacks of pancake coils (section \ref{s.expmeth}). Afterwards, we present and discuss the results for the field dependence and the anisotropy of the critical current of the tape (section \ref{s.IcBath}), the critical current of the coils (section \ref{s.measIccoil}) and the AC loss (section \ref{s.measloss}).


\subsection{Experimental method}
\label{s.expmeth}

\begin{figure}[tbp]
\centering
\includegraphics[width=12 cm]{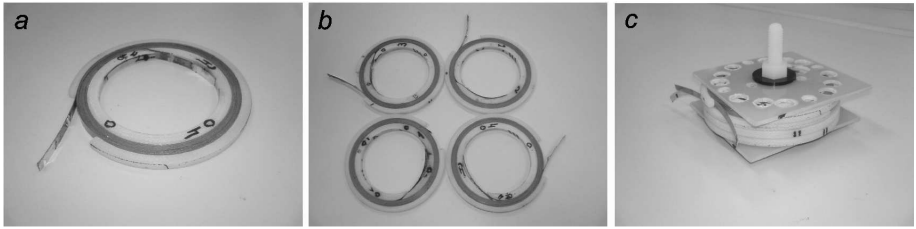}
\caption{\small  The mechanical structure made of epoxy-fiberglass (or fiberglass) avoids vibrations and ensures mechanical stability for both the single pancakes (a,b) and the stacks of pancakes (c) in the experiments.}
\label{f.pictures}
\end{figure}

The pancakes are wound with YBCO coated conductor (CC) tape of width 4.0 mm and self-field critical current $I_c$ = 128 A for a straigh tape at 77 K (tape SCS4050 delivered by SuperPower Inc. \cite{superpower}). Before winding, we insulate the CC tape by means of one layer of Kapton tape of thickness 70 $\mu$m, placing it on the substrate side of the CC tape. The insulated CC tape is then wound on a fiberglass ring with the YBCO layer oriented inward. After winding, the turns of the pancake are fixed with another fiberglass ring surrounding the pancake (figure \ref{f.pictures}). Table 1 lists the basic characteristics of one pancake. The pancakes are assembled to a stack using auxiliary sticks made of thin copper. These sticks are introduced in holes drilled to the fiberglass rings in the direction parallel to the coil axis, crossing the whole pancake height. This ensures a fixed position of the pancakes during the manipulation with the stack. The pancakes are insulated from each other by a thin fiberglass foil. The preparation of the stack is completed by pressing the pancakes. We do this by placing two fiberglass plates on the top and bottom of the stack and tightening them by a non-metallic screw (figure \ref{f.pictures}). Basic properties of the stack are listed in table 1. The pancakes in the stack are electrically connected by soldering their appropriate ends with the superconducting layer face-to-face. We soldered two different kinds of voltage taps, with its corresponding wiring. The first kind are voltage taps at the ends of the coil to measure the whole coil signal. The second ones, are voltage taps at the connections between pancakes of the stack to measure the voltage drop at the connections. All these connections are soldered at around 10 mm from the coil terminals or the connections between pancakes to ensure a good current transfer to the superconducting layer. 

\begin{table}
\caption{\small  \label{t.dimcoils} Dimensions of the coils.}
\vspace{3 mm}
\footnotesize\rm
\begin{tabular*}{\textwidth}{@{}l*{15}{@{\extracolsep{0pt plus12pt}}l}}
{\bf Individual pancakes}\\
\hline
Inner diameter & 60 mm\\
Outer diameter & 67.8 mm\\
Number of turns & 24\\
Tape length & 4.9 m\\
\hline
{\bf Stacks of pancakes}\\
\hline
Insulation thickness between pancakes & 0.3 mm\\
Stack height for: & \\
$\quad$ 2 pancakes & 8.9 mm \\
$\quad$ 3 pancakes & 13.1 mm \\
$\quad$ 4 pancakes & 17.6 mm \\
\hline
\end{tabular*}
\end{table}


We performed the following electrical characterisation of the pancakes and stacks:

\begin{enumerate}
\item $I_c$ was measured for every single pancake (labelled as P1, P2, P3 and P4) using the standard 4-probe method. This monitors the voltage drop across the whole tape length. As criterion for $I_c$ determination, we chose 0.1 $\mu$V/cm instead of the more usual 1 $\mu$V/cm to decrease the possibility of pancake damage.

\item The AC loss was measured for every single pancake using the lock-in method at two frequencies: 23 and 36 Hz. The voltage signal is taken from the taps at the terminals. The transport AC current is measured by a Rogowski coil. We also use this voltage (shifted 90$^{\rm o}$ with respect to the transport current) to compensate the huge inductive component of the measured voltage of the pancake. We set the desired value of this compensation signal by means of a Dewetron DAQP Bridge-B amplifier.
\item Afterwards, we assembled a double pancake (labelled as S2) consisting of pancakes P1 and P2. For these pancakes, we measured the critical current $I_c$ of the individual pancakes forming the stack. For this purpose, we used the appropriate combination of wiring from the voltage taps at at the terminals of the stack and those to monitor the voltage drop at the connections between pancakes.

\item We measured the total AC loss of the double pancake (loss of two assembled pancakes plus dissipation on the soldered contact) using the same approach like in point (ii). The contribution of the soldered contact was also registered for comparison to the total loss.

\item The points (iii) and (iv) were repeated for stacks of pancakes consisting of pancakes P1, P2 and P3 (stack labelled as S3) and all four pancakes (stack labelled as S4).
\end{enumerate}

Our simulations (section \ref{s.sim}) require measurements of the DC properties of the tape used for the pancake preparation. These measurements are the $I_c$ anisotropy in dependence on DC applied magnetic field $B_{\rm DC}$ up to 200 mT and on its direction to the sample surface expressed by angle $\theta$ (sketch in figure \ref{f.IcBath}). For $I_c$ determination, we took the criterion of 1 $\mu$V/cm. The measuring technique is detailed in \cite{coatedIc}.


\subsection{$I_c$ anisotropy and field dependence of the coated conductor}
\label{s.IcBath}

\begin{figure}[tbp]
\centering
\includegraphics[width=8cm]{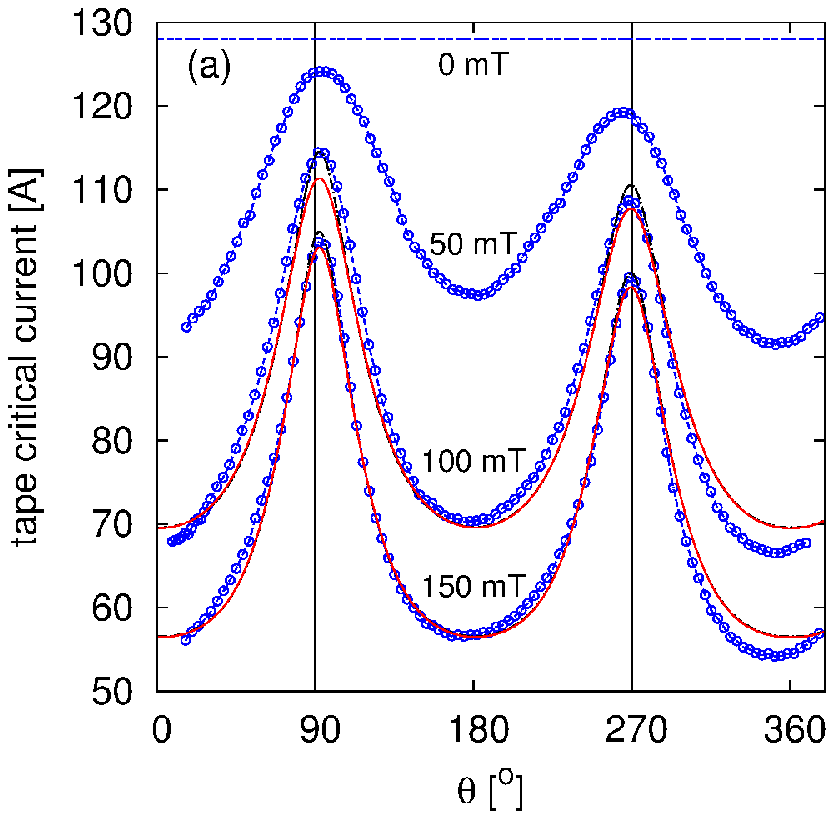}
\includegraphics[width=8cm]{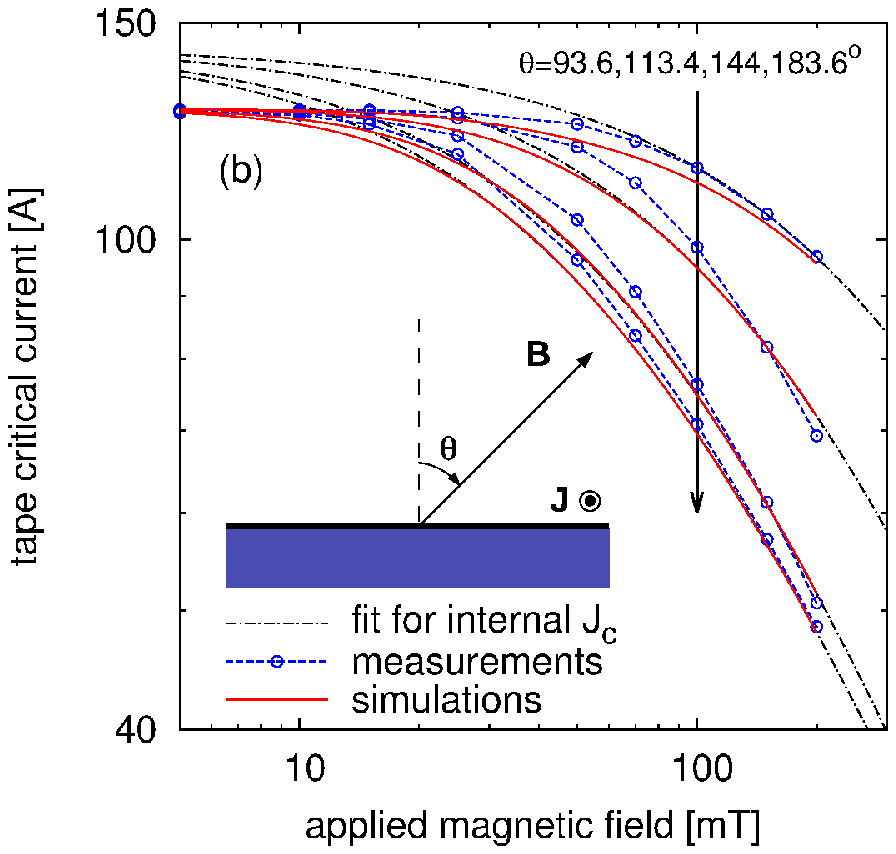}
\caption{\small The (black) chain lines in both graphs are directly calculated from the fit of $J_c$, equations (\ref{Jcall})-(\ref{fpi}), therefore that line represents the critical current if the self-field would not be present. The simulations, (red) solid lines, use the same $J_c$ but taking into account the self-field. (a) The anisotropy of the critical current does not follow a 180$^{\rm o}$ periodicity in the orientation angle $\theta$ [the sketch in (b) shows the $\theta$ definition]. (b) The plateau at low applied fields is caused by the self-field.}
\label{f.IcBath}
\end{figure}

\begin{figure}[tbp]
\centering
\includegraphics[width=8cm]{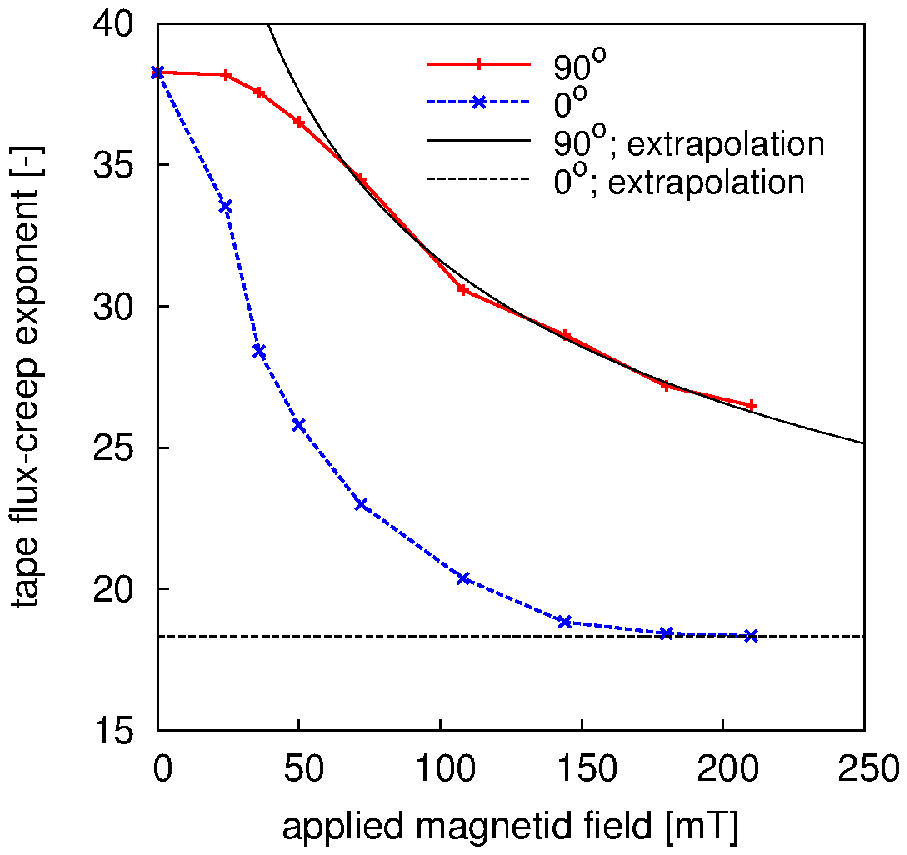}
\caption{\small Measured flux-creep exponent, $n$, from the power-law current-voltage relation of the tape, $I(V)=V_c(I/I_c)^n$. In order to extrapolate to large fields, we can assume that $n$ saturates for $\theta=0^{\rm o}$ (black dash line), while for $\theta=90^{\rm o}$ it fits to $n=(B_{0n}/B)^{1/4}$ with $B_{0n}=1.00\cdot 10^5$ T (continuous black line).}
\label{f.nBa}
\end{figure}

Figure 2a shows $I_c$ of a short sample as a function of the orientation of the applied DC magnetic field $\theta$. The field is oriented perpendicularly to the sample wide surface at position $\theta=0^{\rm o}$ (figure \ref{f.IcBath}b). At position $\theta$ = 90$^{\rm o}$ the vector product ${\bf J}\times {\bf B}_{\rm DC}$ is directed towards the sample surface and at $\theta = 270^{\rm o}$ it is directed towards the YBCO layer-substrate interface. Therefore, the difference between the maxima at $\theta = 90^{\rm o}$ and $\theta = 270^{\rm o}$ can be explained by a stronger surface barrier on the superconductor-air interface \footnote{This interface is afterwards covered with a cap stabilisation layer.} than on the superconductor-substrate one \cite{harrington09APL}. The maximum $I_c$ is not exactly at $\theta = 90^{\rm o}$ and $\theta = 270^{\rm o}$. A similar behaviour was found also in \cite{maiorov05APL,zhangY09PhC} and explained by the presence of additional pinning centres, which was confirmed by SEM analysis. The misalignment of the YBCO layer with the surface also contributes to this shift of the peaks. These irregularities could influence the AC loss of YBCO pancake stacks. For simulation, this influence is taken into account and the $I_c(B_{\rm DC})$ dependence at different $\theta$ is measured and fitted (figure 2b). We discuss more details on the anisotropy in section \ref{s.Jcformula}.

In order to estimate the critical current of the coils for voltage criteria different than that used for the critical current of the tape (1$\mu$V/cm), we measured the current-voltage characteristics, $I(V)$, of the tape at several applied fields and two orientations, 0$^{\rm o}$ and 90$^{\rm o}$. These characteristics present a power-law dependence for all the measured voltage range, from 0.5 to 32.5 $\mu$V/cm. Figure \ref{f.nBa} shows the flux-creep exponent, $n$, of this dependence [$I(V)=V_c(I/I_c)^n$, where $V_c$ is the voltage-per-length criterion for $I_c$]. The $n$ exponent decreases with the applied field and it is different for both measured orientations. For perpendicular applied fields, $n$ is the lowest and it saturates for large applied fields to a relatively large value, 18.3. For parallel applied fields with large enough magnitudes (above 72 mT), the exponent fits well to $n(B,\theta=90^{\rm o})=(B_{0n}/B)^{1/4}$ with $B_{0n}=1.00\cdot 10^5$ T.


\subsection{Critical current of the coils}
\label{s.measIccoil}

We measured the critical currents of the pancakes, both as a standalone coil and forming part of the stacks (tables \ref{t.Icpanc} and \ref{t.Icpancstack}). The critical current of the pancakes forming a stack provide information on the critical current of the complete stack. This is because the critical current of the stack is dominated by the pancake with the lowest critical current. Moreover, by measuring only the global critical current of the stack we could damage the weakest pancake, where the voltage drop of all the coil is concentrated.

There is a very low spread of the critical current of the individual pancakes (0.9A), confirming the homogeneity of the YBCO tape (table \ref{t.Icpanc}). The critical currents of pancakes P1 and P2 forming stack S2 decrease roughly 3 A and are again comparable to each other. A different situation is observed for stack S3. $I_c$ for pancake P3 (placed at the top of stack) in S3 decreases to 54 A. At current $I=55.1$ A, the voltage measured on P3 reaches 500 $\mu$V/cm, whereas the voltage measured on P1 and P2 remains at the background level. To avoid destroying P3, we did not further increase the current, concluding that $I_c$ for P1 and P2 is larger than $55.1$ A (table 3). For the stack S4, the critical current was measured for P1 and P4 (pancakes placed on top and bottom of stack), assuming that their critical current is lower comparing to the inner pancakes. This is justified because the magnetic field in the external pancakes is larger than in the inner ones. Moreover, simulations show that the pancakes with the lowest critical current are the top and bottom ones (section \ref{s.Iccoils}). Surprisingly, P1 and P4 in S4 had higher $I_c$, 57.4 A, than P3 in stack S3 (table 3). A possible explanation is an inhomogeneity of $J_c$ along the tape length that appears at relatively large magnetic fields, around 200 $\mu$T. In order to check that pancake P3 was not damaged during the stack manipulation, P3 was removed from the stack and measured again, showing exactly the same critical current as before assembling the stack. In general, the critical current of the coils decreases with the number of pancakes because the local self-field increases.

Stack S3 (or the weakest pancake in stack S3) has around 10\% lower critical current comparing to the interpolation line between the values for stack S2 and S4 (figure \ref{f.Icnp}). This 10\% difference can be attributed to the non-homogeneity in the tape length. Actually, the homogeneity of the tape in reel-to-reel tests is usually based on critical current measurements in self-field. Then, it is not surprising that the in-field critical current of the tape has a larger variation than the rated 5\% from the producer. This larger variation of the in-field critical current will inevitably appear in coils made of state-of-the-art tapes, and thence it has to be taken into account in the coil design.

\begin{table}
\caption{\small \label{t.Icpanc} Critical current of the individual pancakes.}
\vspace{3 mm}
\footnotesize\rm
\begin{tabular*}{\textwidth}{@{}l*{15}{@{\extracolsep{0pt plus12pt}}l}}
{\bf pancake label} & {\bf P1} & {\bf P2} & {\bf P3} & {\bf P4} \\
\hline
$I_c$ [A] & 68.5 & 68.0 & 68.3 & 67.6\\
\end{tabular*}
\end{table}

\begin{table}
\caption{\small \label{t.Icpancstack} Critical current of each pancake within the stacks S2, S3 and S4 made of 2, 3 and 4 pancakes, respectively. The stacks are constructed from the pancakes P1, P2, P3 and P4, which critical current as standalone coil is in table \ref{t.Icpanc}. The dimensions of the pancakes and the stacks are in table \ref{t.dimcoils}.}
\footnotesize\rm
\begin{tabular*}{\textwidth}{@{}l*{15}{@{\extracolsep{0pt plus12pt}}l}}
\hline
{stack/pancake label} & {critical current [A]}\\
\hline
 {\bf S2} & \\
 P1 &	65.5 \\
 P2 &	65.2 \\
\hline
 {\bf S3 } & \\
P1 &$>$55.1 \\
P2 &$>$55.1 \\
P3 &54.0 \\
\hline
{\bf S4 } & \\
P1 &	57.4  \\
P2 &	-- \\
P3 &	-- \\
P4 &	57.4 \\
\hline
\end{tabular*}
\end{table}


\subsection{AC loss}
\label{s.measloss}

\begin{figure}[tbp]
\centering
\includegraphics[width=8 cm]{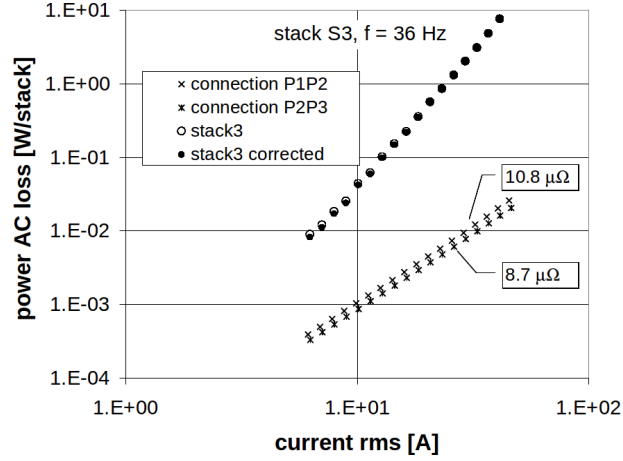}
\caption{\small Measured power AC loss of the 3-pancake stack, S3, before (open circles) and after (full circles) correcting the contribution of the soldered connections. This contribution is much smaller the total loss (crosses and asterisks are for the connections between pancakes P1,P2 and P2,P3, respectively) because of the relatively low connection resistance (labels in the graph).}
\label{f.QRexp}
\end{figure}

\begin{figure}[tbp]
\centering
\includegraphics[width=8 cm]{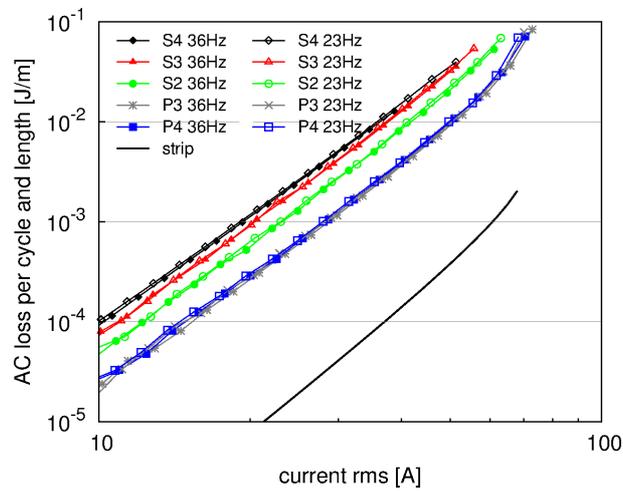}
\caption{\small The measured AC loss per cycle and unit tape length for two frequencies, 36 and 23 Hz, shows that the AC loss in the coils is of hysteresis nature. The AC loss per tape length in the stacks of 2, 3 and 4 pancakes (S2, S3, S4) is larger than that for the individual pancakes (only pancakes P3 and P4 from table \ref{t.Icpanc} are shown) and around 2 orders of magnitude larger than a single tape, represented by the Norris' strip formula \cite{norris70JPD}.}
\label{f.Qexp}
\end{figure}

The measured resistance of the soldered contacts between pancakes P1P2, P2P3 and P3P4 were 10.9, 8.8 and 5.7 $\mu\Omega$, respectively. The influence of these contacts on the total power AC loss of the stacks of pancakes is negligible. As illustration, figure \ref{f.QRexp} shows the dissipation on the contacts between pancakes P1, P2 and P2, P3 together with the total one of stack S3, both before and after subtracting the dissipation on the contacts. The contribution from the soldered contacts is only visible at the lowest transport current.

After removing the contacts contribution, the AC loss per cycle and unit tape length presents the following features (see figure \ref{f.Qexp} for the current frequencies 23 and 36 Hz). The AC loss\footnote{In order to avoid repetition, we refer to the AC loss per cycle by simply ``AC loss". With the exception of figure \ref{f.QRexp}, all the results in this article are for the AC loss per cycle and unit tape length.} for all the individual pancakes is the same within the measurement accuracy (for simplicity, figure \ref{f.Qexp} shows the results for P3 and P4 but not P1 and P2). It presents no frequency dependence, confirming the pure hysteresis nature assumed in the simulations. The AC loss per tape length increases with increasing the number of pancakes. This means that the stack has higher AC loss than the sum of the ac loss in its pancakes when forming a standalone coil. In particular, the AC loss per tape length is around 3 and 4 times larger than that of a single pancake for the 2- and 4-pancake stacks, respectively. Moreover, the AC loss per tape length for the 4-pancake stack is around two orders of magnitude larger than for a single tape. This loss is much larger than for the two AC applications closest to commercialisation, power transmission cables and resistive fault-current limiters, where the AC loss per tape length is below that of one tape. In order to compare the results with usual desired values in transformers, we take the requirement of the AC loss per unit tape length in \cite{pleva10SST} for a 25 MVA self-limiting transformer: 44 mW/m. The AC loss per unit tape length in the measured 4-pancake stack is around 400 mW/m at a current amplitude of 85\% of the critical current (49 A) and 50 Hz of frequency. This suggests that the AC loss in close-packed stacks of pancake coils, like in the experiment, will be too high for many transformer designs.


\section{Simulations and comparison to experiments}
\label{s.sim}

Simulations can only be useful for the design of a coil if they provide a reliable prediction {\em before} constructing the coil, not only an interpretation of the measurements after building the coil. The simulations in this section achieve this goal, regarding the electrical properties. For this, it is necessary to first know the magnetic field and angular dependence of $J_c$, $J_c(B,\theta)$, of the coated conductor and the coil dimensions. Below, we first present the simulation technique (section \ref{s.simtech}), then we extract the field and angular dependence of $J_c$ for the coated conductor (section \ref{s.Jcformula}) and finally we present the results: the magnetic field and current distribution (section \ref{s.BJ}), the coils critical current (section \ref{s.Iccoils}) and the AC loss (section \ref{s.losscalc}). In that section, we also discuss the effect on the AC loss by adding more pancakes to our stack and the AC loss reduction by a 2-strand Roebel cable.


\subsection{Simulation technique}
\label{s.simtech}

First, we extract the field, $B$, and angle, $\theta$, dependence of the internal $J_c$ of the coated conductor from the measurements of the critical current, $I_c$, as a function of the {\em applied} field and its orientation (figure \ref{f.IcBath}). This extraction is not straightforward because the self-field of the tape influences the local $B$ and $\theta$, and thence $J_c$. We obtain $J_c(B,\theta)$ from $I_c$ measurements with the method we presented in \cite{coatedIc}. Outlining, we developed a computer program that calculates the critical current of a tape, given a certain $J_c(B,\theta)$ function. This program is based on the self-consistency of the critical current density and the local magnetic field, taking into account that the current distribution creates magnetic field. With this program, we fit the parameters of a function for $J_c(B,\theta)$ in such a way that the measured in-field anisotropy of $I_c$ agrees the best with the computations. The choice of the function of $J_c(B,\theta)$ is based on physical arguments. For a single tape with a field-dependent $J_c$, it is not possible to assume that the electromagnetic quantities are uniform across the thickness. This is because (i) the parallel magnetic field influences $J_c$ and (ii) the parallel magnetic field is antisymmetric with respect to the tape thickness, so the average in the thickness vanish but not its local value. Indeed, the parallel field at the tape surface is of the same order of magnitude as the perpendicular magnetic field. For the computations, we divide the cross-section of the superconducting layer (1.4 $\mu$m $\times$ 3.96 mm) in 17$\times$170 rectangles with uniform $J_c$. In the rest of the article we refer to the superconducting layers as simply ``tapes".

Once the $J_c(B,\theta)$ dependence is known, we calculate the AC loss by means of the Minimum Magnetic Energy Variation (MMEV) method. The MMEV method assumes the sharp $E(J)$ relation of the critical-state model, where $E$ is the electrical field. It is based on, first, a variational principle proposed by L. Prigozhin \cite{prigozhin97IES}, which finds the current distribution by minimising the magnetic energy variation, and, second, a fast non-standard minimisation routine. This routine has been developed incrementally in several articles. Initially, Sanchez and Navau solved a cylinder in an applied magnetic field under certain restrictions \cite{sanchez01PRB}. Later on, Pardo {\it et al.} developed the general method for tapes under any combination of applied magnetic field and transport current \cite{HacIacinphase}. The latest stage of the method is published in \cite{pancakeBi,pancakeFM}, where \cite{pancakeBi} and \cite{pancakeFM} take into account the field dependence of $J_c$ and the interaction with linear magnetic materials, respectively. Independently, L. Prigozhin and V. Sokolovsky applied the variational principle in \cite{prigozhin97IES} to solve single pancake coils by more standard minimisation techniques \cite{prigozhin11SST}. The MMEV method is also useful to calculate the critical current of superconducting coils \cite{pancakeBi,coatedIc}. The main advantage of MMEV compared to conventional Finite-Element Methods is its speed \cite{roebelcomp} and low RAM memory. Thanks to this, it is possible to calculate the AC loss in coils made of many turns \cite{pancaketheo}. In this article, we report results for coils (in this case stacks of pancake coils) with as many as $24\times32=$768 turns (figure \ref{f.Qmany}).

The MMEV method is also useful to calculate the critical current of the coils for a certain $J_c(B,\theta)$. The critical current is calculated by solving the current distribution for an increasing transport current from the zero-field cooled state. The calculation stops when there is at least one turn that is fully penetrated by critical current, corresponding to the coil critical current. The details of the MMEV method for a $J_c(B,\theta)$ dependence can be found in \cite{pancakeBi}.

For the coils, we assume that the electromagnetic quantities are uniform across the tape thickness. This is a good approximation for pancake coils because the variation of the parallel component of the magnetic field across the thickness is much smaller than the average value, contrary to single tapes. In this article, we use between 120 and 76 current elements across the width in each turn and one element in their thickness. We only use the lowest number of elements for coils with more than 100 turns. 

We calculate the magnetic flux lines (figure \ref{f.rA}) as level curves of $rA$ \cite{brandt98PRBa}, where $r$ is the radius and $A$ is the vector potential in the Coulomb's gauge. This gives the exact direction of the magnetic flux density (or magnetic field), ${\bf B}=\mu_0{\bf H}$. However, the separation of these level curves is not inversely proportional to $B$ but to $rB$. For our case, the coils thickness is one order of magnitude smaller than the average radius. Then, the separation between level curves of $rA$ approximately describes the intensity of the local flux density.

We use the experimental tape and coil dimensions for the calculations. In particular, the dimensions of the superconducting layer of the tapes is 1.4 $\mu$m $\times$ 3.96 mm, the separation between the superconducting layers of neighbouring turns in the pancakes is 188 $\mu$m, their inner radius is 29.5 mm and the superconductor-to-superconductor gap between pancakes is 540, 573, 465 $\mu$m for the 2-, 3- and 4-pancake stacks respectively. This gap is 465 $\mu$m for all the stacks of more than 4 pancakes and for all the stacks calculated in figure \ref{f.Qmany}.


\subsection{Formula for the anisotropic magnetic-field dependence of $J_c$}
\label{s.Jcformula}

Next, we present the formula for the anisotropic field dependence of the internal critical current density, $J_c$. The critical current density for which the computed critical current, $I_c$, (see section \ref{s.simtech}) fits the best to the measurements (figure \ref{f.IcBath}) is 
\begin{eqnarray}
J_c(B,\theta,J) & = & [ J_{c,ab}(B,\theta,J)^m+J_{c,c}(B)^m ]^{1/m} \label{Jcall}
\end{eqnarray}
with
\begin{eqnarray}
J_{c,ab}(B,\theta,J) & = & \frac{J_{0,ab}}{\left[{1+\frac{Bf(\theta,J)}{B_{0,ab}}}\right]^{\beta_{ab}}}, \label{Jcab} \\
J_{c,c}(B) & = & \frac{J_{0,c}}{\left[{1+\frac{B}{B_{0,c}}}\right]^{\beta_{c}}}, \label{Jc} \\
\end{eqnarray}
and
\begin{eqnarray}
f(\theta,J)= & = & \left \{ \begin{array}{ll} 
                              f_{0}(\theta) & {\rm if}\ J\sin\theta > 0 \\
															f_{\pi}(\theta) & {\rm otherwise} 
															\end{array} \right . , \label{f} \\
f_{0}(\theta) & = & \sqrt{u^2\cos^2(\theta+\delta_{0})+\sin^2(\theta+\delta_{0})}, \label{f0} \\
f_{\pi}(\theta) & = & \sqrt{u^2\cos^2(\theta+\delta_{\pi})+v^2\sin^2(\theta+\delta_{\pi})}, \label{fpi}
\end{eqnarray}
where the parameters are $m=8$, $J_{0,ab}=2.53\cdot 10^{10}$ A/m$^2$, $J_{0,c}=2.10\cdot 10^{10}$ A/m$^2$, $B_{0,ab}=414$ mT, $B_{0,c}=90$ mT, $\beta_{ab}=0.934$, $\beta_c=0.8$, $u=5.5$, $v=$1.2, $\delta_0$=-2.5$^{\rm o}$ and $\delta_\pi$=0.5$^{\rm o}$. The critical current density at zero local field is $J_c(B=0)=2.59\cdot 10^{10}$ A/m$^2$.

The critical current calculated with this formula agrees with the experiments (figure \ref{f.IcBath}). The maximum deviation is around 5\% at intermediate applied fields (between 20 and 100 mT), although the average deviation is below 4\%. As discussed in \cite{coatedIc,hengstberger10APL,sanchez10APL}, the local $J_c$ at a zero local field is larger than the average critical current density at zero applied field, that is at self-field (figure \ref{f.IcBath}b). However, for the tape in this article, this effect is less important than in \cite{coatedIc}.

The formula for $J_c$ [equations (\ref{Jcall})-(\ref{fpi})] is similar to that in \cite{coatedIc}. In particular, $J_c$ has several contributions with a field dependence of the Kim type \cite{kim62PRL} and elliptical anisotropy. However, there are some differences with the formula in \cite{coatedIc}. The most important features of the formula in this article are the following.

First, the sign of the current density, $J$, influences $J_c$ through equation (\ref{f}). This is in order to describe the surface barrier effect, responsible for the difference in the height of the peaks close to the $ab$ direction (90$^{\rm o}$ and 180$^{\rm o}$ in figure \ref{f.IcBath}) \cite{harrington09APL}. For our case, $J_c$ is lower when the driving force is towards the interface between the superconductor and the buffer layers, consistent with our measurements (figure \ref{f.IcBath}) and those in \cite{harrington09APL}.

Second, the anisotropic contribution, $J_{c,ab}$ in equation (\ref{Jcall}), is due to isotropic pinning of the elliptical anisotropic vortices. This is supported by: (i) the angular dependence is described by an elliptical function [equations (\ref{f0}),(\ref{fpi})] and (ii) the experimental anisotropy factor, $u$=5.5, is close to the mass anisotropy of YBCO, 5.3 \cite{blatter94RMP,durrell04PRB}. The deviation of the $ab$ peak from the $ab$ direction is because of the tilt in the $ab$ planes \cite{maiorov05APL,zhangY09PhC}. Moreover, close to the $ab$ directions the vortex should deform, resulting in a complicated 3D pinning scenario \cite{durrell04PRB}. Possible artificial pinning centres close to the $ab$ direction further complicate the pinning landscape \cite{zhangY09PhC}. The shift in the $ab$ peaks close to 90$^{\rm o}$ and 270$^{\rm o}$ is different to each other: $\delta_0$=-2.5$^{\rm o}$ and $\delta_\pi$=0.5$^{\rm o}$. This difference could be explained by the experimental error in setting the angle origin, around 1$^{\rm o}$.

Finally, the isotropic contribution, $J_{c,c}$ in equation (\ref{Jcall}), is due to anisotropic pinning in the $c$ direction caused by columnar defects, such as dislocations and nano-particle arrays. Indeed, an anisotropic pinning can result in an apparent isotropic contribution. This is because the angular dependence of the critical current density associated to these defects has a wide peak around the $c$ direction (0$^{\rm o}$ and 180$^{\rm o}$). Since in our case this contribution is only dominant at a small range close to the $c$ direction, an isotropic or anisotropic contribution results in basically the same total $J_c$. We expect that a wide peak around the $c$ direction will appear in large applied fields, as in \cite{maiorov05APL,zhangY09PhC,selvamanickam09PhC}. The lower decay exponent $\beta$ in the $c$ direction supports this prediction.


\subsection{Field and current distribution}
\label{s.BJ}

\begin{figure}[tbp]
\centering
\includegraphics[width=9 cm]{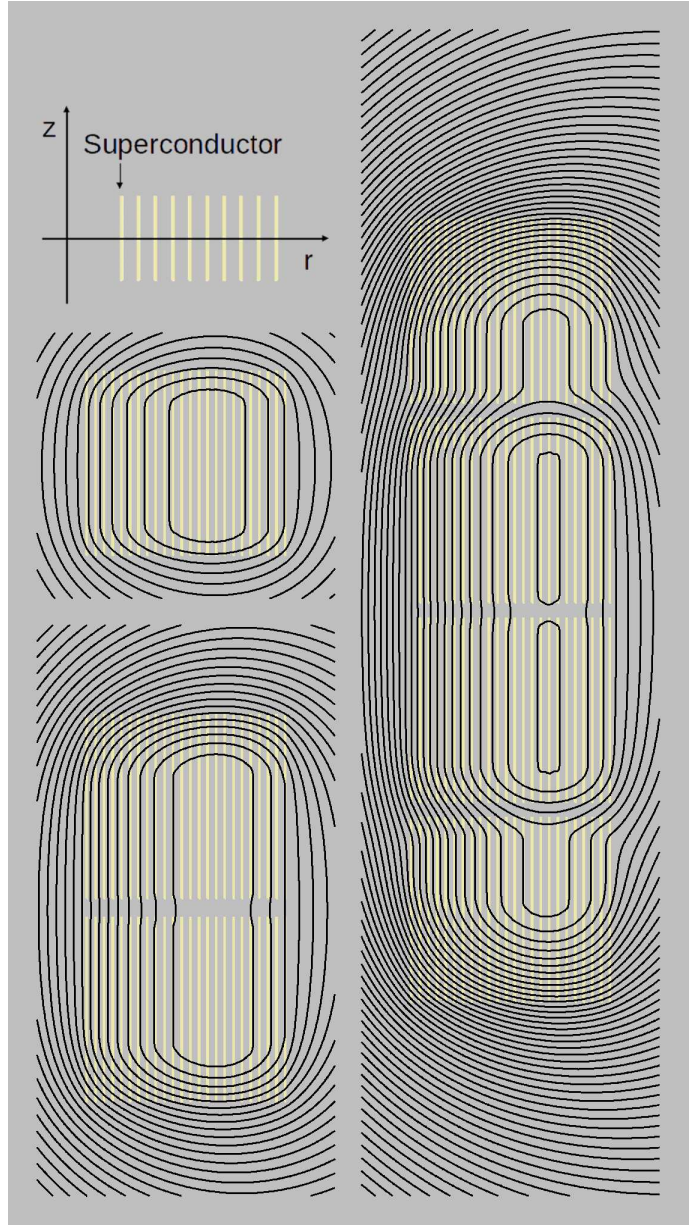}
\caption{\small Cross-section of the coils consisting of 1, 2 and 4 pancakes with the computed flux lines (black lines) at the peak of the AC cycle for 46.0 A of amplitude; see sections \ref{s.simtech} and \ref{s.Jcformula} for the geometrical parameters and the $J_c(B,\theta)$ dependence, respectively. The scketch defines the axial and radial coordinates, $z$ and $r$, respectively. The flux lines show that a double pancake mainly behaves as a single one with double width but a stack of more pancakes, like 4, is very different. These flux lines are calculated as level curves of $rA$ (section \ref{s.simtech}). All the graphs are at the same scale, including the difference in $rA$ between neighbouring flux lines. However, the thickness of the superconducting layer (light yellow lines) is exaggerated for visibility.}
\label{f.rA}
\end{figure}

\begin{figure}[tbp]
\centering
\includegraphics[width=7.78 cm]{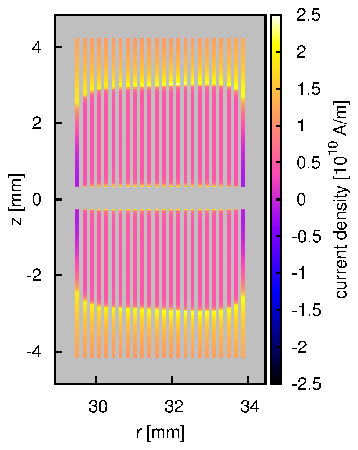}
\caption{\small This cross-section shows the current distribution in a double pancake at the peak of the AC cycle for 46.0 A of amplitude (same situation as figure \ref{f.rA}). For better visualisation, the colour-graded lines representing the superconducting layer are shown with a larger thickness than the one used in the simulations. The represented quantity is actually the average current density over the thickness of the tapes.}
\label{f.J24x2}
\end{figure}

\begin{figure}[tbp]
\centering
\includegraphics[width=7.78 cm]{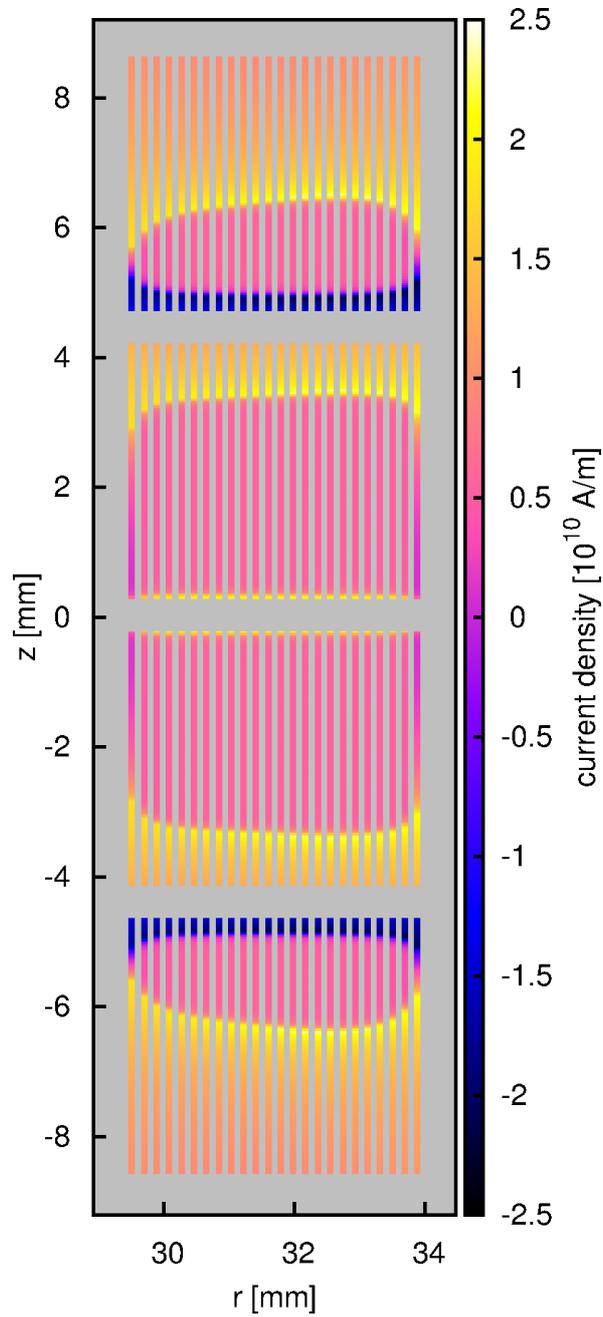}
\caption{\small This cross-section shows the presence of current density with opposite sign from the transport current at the top and bottom pancakes for the 4-pancake stack. The calculated situation is at the peak of the AC cycle for 46.0 A of amplitude (same situation as figure \ref{f.rA}). For better visualisation, the colour-graded lines representing the superconducting layer are shown with a larger thickness than the one used in the simulations. The represented quantity is actually the average current density over the thickness of the tapes.}
\label{f.J24x4}
\end{figure}

\begin{figure}[tbp]
\centering
\includegraphics[width=10 cm]{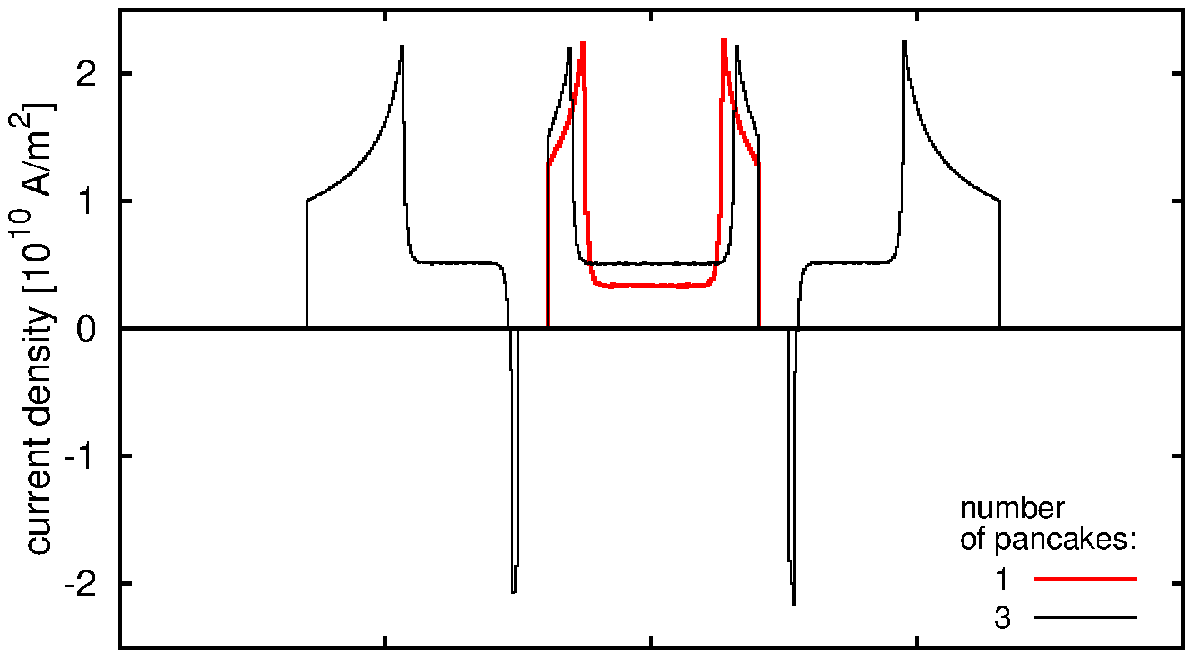}
\includegraphics[width=10 cm]{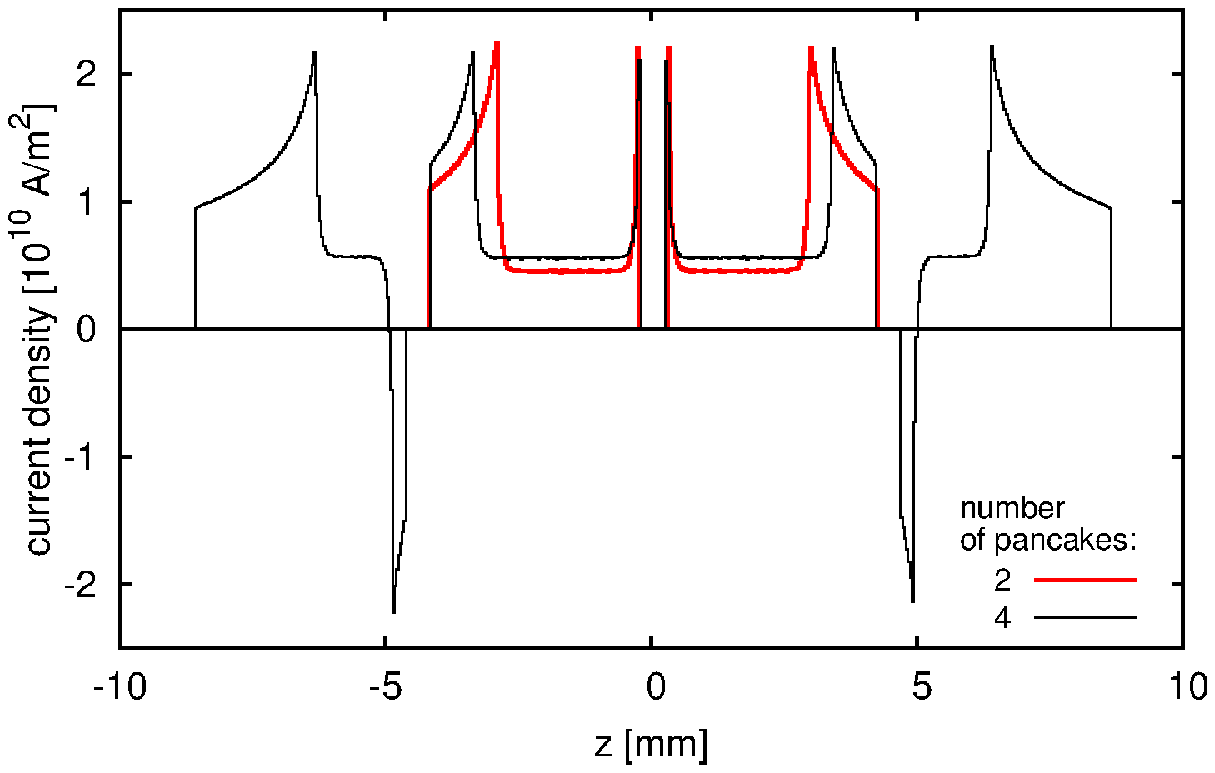}
\caption{\small Computed current density in the central turns (12th turn counting from the one at the inner radius) at the peak of the AC cycle for 46.0 A of amplitude (same situation as figure \ref{f.rA}). Uniform current distribution would result in 0.83$\cdot 10^{10}$ A/m$^2$. The vertical position, $z$, is defined in the caption of figure \ref{f.rA}. The 3- and 4-pancake stacks present currents with opposite sign as the transport current but not the single and double pancakes. The represented quantity is actually the average current density over the thickness of the tapes.}
\label{f.Jcen}
\end{figure}

\begin{figure}[tbp]
\centering
\includegraphics[width=7.78 cm]{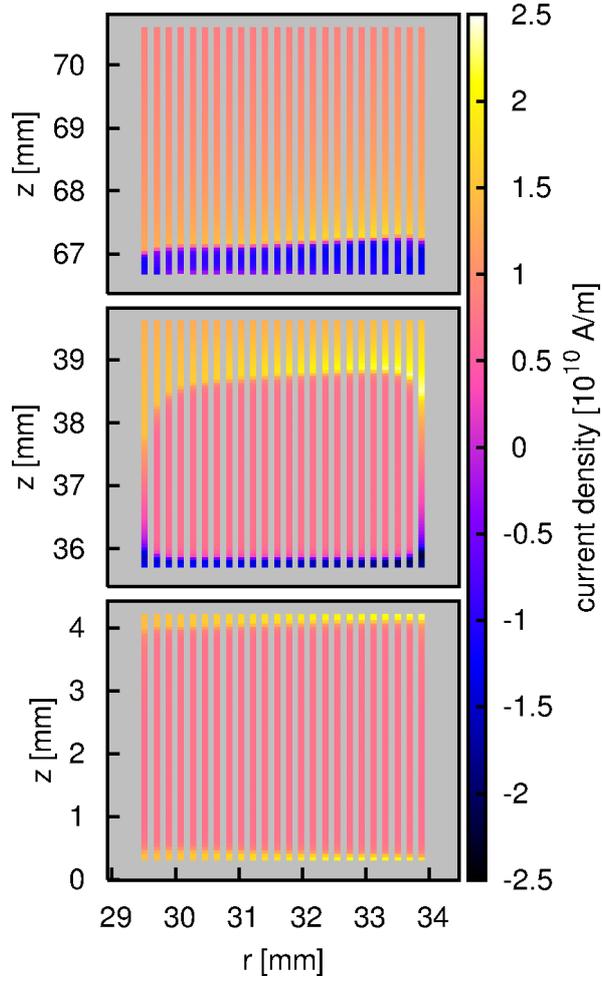}
\caption{\small These cross-sections show the current distribution in the pancakes at the top, 3/4 height and central pancakes (1st, 8th and 16th pancakes from the top) for a 32-pancake coil, respectively, from top to bottom. The situation is for the peak of the AC cycle for 46.0 A of amplitude. See sections \ref{s.simtech} and \ref{s.Jcformula} for the geometrical parameters and the $J_c(B,\theta)$ dependence, respectively. For better visualisation, the colour-graded lines representing the superconducting layer are shown with a larger thickness than the one used in the simulations. The represented quantity is actually the average current density over the thickness of the tapes.}
\label{f.J24x32}
\end{figure}

\begin{figure}[tbp]
\centering
\includegraphics[width=12 cm]{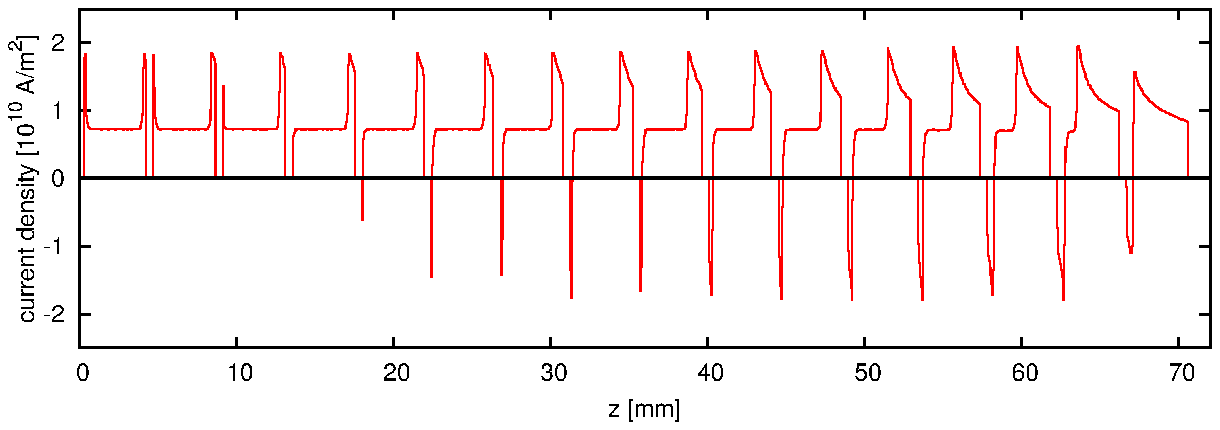}
\caption{\small Computed current density in the central tapes (12th tape from the inner radius) for a stack of 32 pancakes and 24 turns per pancake (only the upper half is shown). The situation is at the peak of the AC cycle for 46.0 A of amplitude (same situation as figure \ref{f.J24x32}). Uniform current distribution would result in 0.83$\cdot 10^{10}$ A/m$^2$. The under-critical region, including the plateau, shrinks with approaching to the end of the stack. The represented quantity is actually the average current density over the thickness of the tapes.}
\label{f.Jcenmany}
\end{figure}

In this section, we first discuss the magnetic field lines (figure \ref{f.rA}), the current distribution (figures \ref{f.J24x2}, \ref{f.J24x4} and \ref{f.J24x32}) and the current density in the central turns, that is the 12th turn from the inner radius, (figures \ref{f.Jcen} and \ref{f.Jcenmany}).

The magnetic field is parallel to the tapes in a large portion of the coil, as expected from single pancake coils \cite{pancaketheo} (figure \ref{f.rA}). This is because the tapes shield the perpendicular component of the magnetic field created by the other tapes. Due to the vertical symmetry, this component of the magnetic field also vanishes at the central gap of the double pancake and the four-pancake coil (and any coil with an even number of pancakes). On the contrary, the perpendicular component does not vanish at the other gaps, such as the top and bottom gaps of the four-pancake coil (figure \ref{f.rA}). The region with zero perpendicular field shrinks with increasing the current until it becomes a single horizontal line at the critical current.

The magnetic field lines and the current distribution show that a double pancake coil roughly behaves as a single pancake with double width (figures \ref{f.rA} and \ref{f.Jcen}). This is because of the vertical symmetry of the coils and the fact that the current can freely distribute accross the upper or lower half for a given radial position, as it is the case for a single pancake. However, the case of a four-pancake coil is qualitatively different. The cause is the presence of current density with opposite sign to the net current at the top and bottom pancakes (see text below). In the following, we detail the main features of the magnetic field and the current distribution.

At the peak of the AC current, the four-pancake stack presents current density with opposite sign from the transport current, which can only be due to magnetisation currents (figure \ref{f.J24x4}). These currents with opposite sign do not appear in a double pancake (figure \ref{f.J24x2}). The cause is the mirror symmetry with respect to the horizontal mid-plane. In more detail, the magnetisation field is antisymmetric. For this situation, the maximum exclusion of the perpendicular field in the superconducting volume for a given transport current is when current density with the same sign as the transport current accumulates at the top and bottom of the coil. This situation is similar to an slab with transport current. For the 4-pancake stack, the current density cannot accumulate at the top and bottom of the whole coil because the net current in the tapes of the central pancakes has to be the same as those in the top and bottom ones. The current tries to accumulate at the top and bottom of the whole coil, creating a large region with positive current at the top and bottom coils, but there appears a negative current density to keep the transport current at the given value (figure \ref{f.J24x4}). This negative current density contributes to the relatively large current penetration in the inner pancakes near to the side neighbouring the top and bottom pancakes. The gap between these pancakes further contributes to the current penetration.

The current distribution and magnetic flux lines for the sub-coil forming the central two pancakes of the 4-pancake stack are qualitatively similar to that of a double pancake (figures \ref{f.rA}-\ref{f.J24x4}). This is because for both cases the magnetic field is anti-symmetric and the current can freely distribute across the half-height of the coil/sub-coil. The main difference is that for a double pancake the current penetration and perpendicular field at the top and bottom are larger. This is due to the fact that for 4 pancakes, the top and bottom pancakes create a perpendicular magnetic field opposing to that from the central sub-coil. A similar situation appears when comparing the central pancake of a 3-pancake stack with a single pancake (only the current distribution at the central pancake is shown in figure \ref{f.Jcen}).

As usual for coils and stacks of tapes, the current density in one pancake is qualitatively the same for all the turns of the same pancake, except at the inner and outer radius of the coil \cite{tapesfull,pancaketheo} (see figure \ref{f.J24x4}). Then, the properties of the current density at the central turn are representative for the complete coil.

In the following, we discuss the details of the current density in the central turn (figures \ref{f.Jcen} and \ref{f.Jcenmany}). Actually, the simulations calculate the average current density across the thickness of the tape.

First, the average current density presents two sharp peaks in each tape (figures \ref{f.Jcen} and \ref{f.Jcenmany}). These peaks separate the critical and sub-critical regions, as for thin strips \cite{brandt05PRBa}. The critical regions are those which the average current density, $J$, becomes the critical current density, $J_c$. The sub-critical ones are those with $|J|<J_c$. The sharp peaks appear because, at the boundary between both regions, the perpendicular field vanishes and $J_c$ are maximum. Between the peak and the tape boundary, the average current density decreases because the perpendicular field increases (figure \ref{f.rA}), decreasing $J_c$. The average current density is uniform in practically all the under-critical region. The cause of this is detailed in \cite{roebelcomp} and is due to the fact that the magnetic flux does not cross the under-critical region and the separation between tapes is much smaller than the tapes width. This average current density times the tape thickness, $Jd$, equals to the difference in the parallel magnetic field between the surfaces of the tape. 

Second, the situation of a double pancake is practically the same as a single pancake with double width (figure \ref{f.Jcen}). The only difference is the presence of the narrow peaks next to the gap, evidencing a slight penetration of critical current density in that region.

Finally, in a stack of many pancakes the under-critical region shrinks with approaching to the end of the coil, disappearing at the top pancake (see figure \ref{f.Jcenmany} for 32 pancakes). The cause is the following. The perpendicular magnetic field increases with approaching to the coil end. This creates magnetisation currents, including negative current density at the peak of the AC cycle. Then, the region with positive critical current density has to expand in order to keep the same net current. As a result, the under-critical region shrinks. Another issue is that the peak of the current density at the uppermost pancake is lower than for the other pancakes. This is because there is no under-critical region, so the perpendicular field is non-zero, increasing $B$ and decreasing $J_c$.

For the 32-pancake coil it is worth to present the current distribution in all the turns at least for 3 of its pancakes: the central one, the top one and the pancake at 3/4 heigh, that is the 8th pancake from the top (figure \ref{f.J24x32}). Apart from the features already visible for the central pancake (figure \ref{f.Jcenmany}), figure \ref{f.J24x32} shows that the penetration depth of the critical region is larger at the inner turns. The cause is that the parallel magnetic field at the inner turns is larger, decreasing the critical current density so that it is necessary a larger penetration depth to achieve the same transport current. The current density in the plateau of the under-critical region is the same for all the turns, except at the inner and outer turns. This is because this current density is proportional to the variation of the parallel magnetic field in the radial direction, which is constant for a long coil. These effects are already visible for 4 pancakes but for 32 pancakes they are more pronounced.


\subsection{Critical current of the coils}
\label{s.Iccoils}

\begin{figure}[tbp]
\centering
\includegraphics[width=7.78 cm]{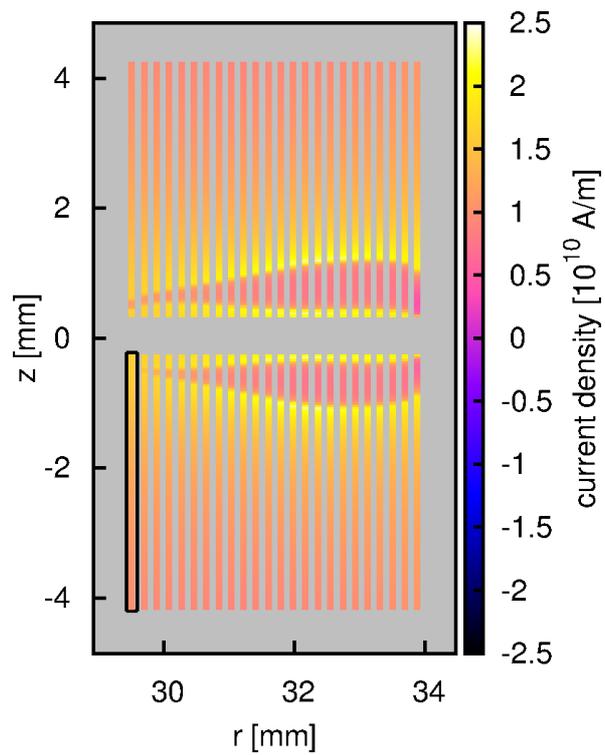}
\caption{\small This cross-section shows the current distribution at the critical current (70 A) for the 2-pancake stack. The black frame highlights the turn limiting the critical current. For better visualisation, the colour-graded lines representing the cross-section of the superconducting layer are shown with a larger thickness than the one used in the simulations. The represented quantity is actually the average current density over the thickness of the tapes.}
\label{f.JIc2}
\end{figure}

\begin{figure}[tbp]
\centering
\includegraphics[width=7.78 cm]{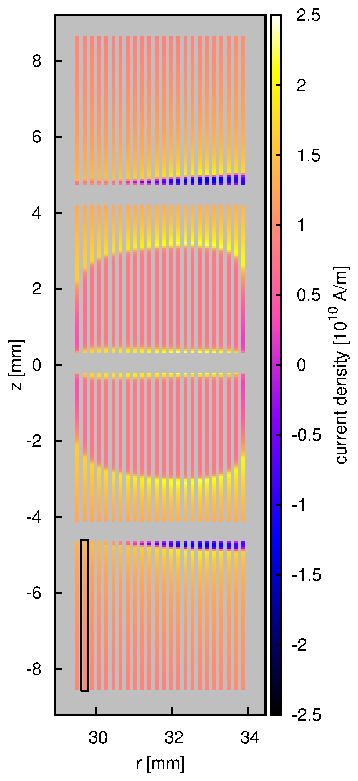}
\caption{\small This cross-section shows the current distribution at the critical current (60 A) for the 4-pancake stack. The black frame highlights the turn limiting the critical current. For better visualisation, the colour-graded lines representing the cross-section of the superconducting layer are shown with a larger thickness than the one used in the simulations. The represented quantity is actually the average current density over the thickness of the tapes.}
\label{f.JIc4}
\end{figure}

The simulations show that the turn that limits the most the critical current for the double pancake is the innermost turn at the bottom of the coil (figure \ref{f.JIc2}). This is because the magnetic field is the largest there and the anisotropy of the tape is not very large (figure \ref{f.IcBath}). This is in contrast to coils made of Bi2223 tapes, where the limiting turn is at the average radius on the top and bottom ends \cite{pitel01SST}. The reason is that Bi2223 tapes are highly anisotropic. In addition, the critical current density in the YBCO tape is asymmetric, so the bottom pancake has slightly smaller critical current than the top one. 

For the stack of 4 pancakes, the turn that limits the most the critical current is the second turn from the inner radius at the bottom of the pancake (figure \ref{f.JIc4}). This is because there is a compromise between the magnitude of the magnetic field (the largest field is at the inner radius) and its orientation (with increasing the radius, the magnetic field becomes more perpendicular to the tape). In contrast to 2 pancakes, for 4 pancakes the anisotropy has a certain small effect because the magnetic field is larger, and thence the anisotropy is more important (see figure \ref{f.IcBath}).

\begin{figure}[tbp]
\centering
\includegraphics[width=8 cm]{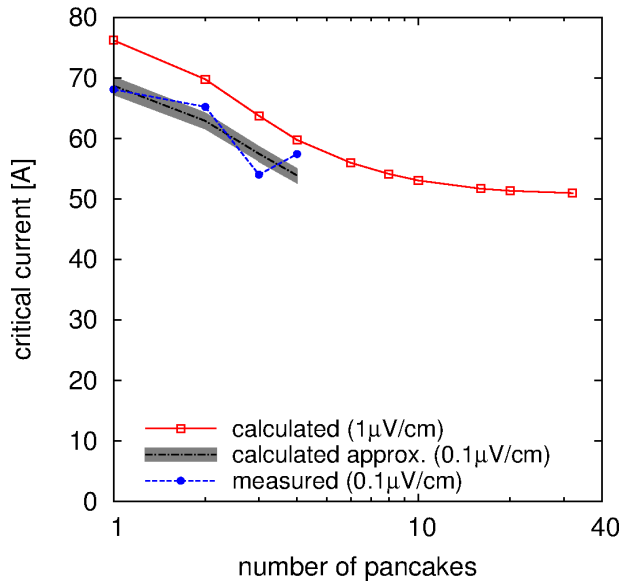}
\caption{\small The simulations of the critical current of the stacks agree with the measurements, with a maximum dispersion of 6.4\% and only 0.9\% difference for one pancake. The quantity in parenthesis is the voltage criterion. The grey area represents the error committed by extrapolating the calculations for 1$\mu$V/cm to to 0.1 $\mu$V/cm. See sections \ref{s.measIccoil} and \ref{s.Iccoils} for the measurements and simulations, respectively. The separation between pancakes for more than 4 pancakes is the same as for 4 pancakes.}
\label{f.Icnp}
\end{figure}

Simulations provide a good estimate of the critical current of the stack of pancake coils (figure \ref{f.Icnp}). The simulations taking $J_c$ from equations (\ref{Jcall})-(\ref{fpi}) over-estimate the critical current. The main reason is the different voltage criterion for the measurements of the critcal current for the coils and the tape: 0.1 and 1 $\mu$V/cm, respectively. The simulations, based on $J_c$ obtained from measurements with a larger voltage criterion, predict a larger critical current. In order to estimate the critical current for 0.1 $\mu$V/cm, we use the measured $n$ exponent of the current-voltage curve of the tape (figure \ref{f.nBa}). This $n$ exponent depends on the magnetic field and its orientation. What we do is to calculate the upper and lower limit of $I_c$ using the upper and lower $n$ in the coils. Simulations at the critical current for 1$\mu$V/cm show that the maximum magnetic field at the turn that limits the critical current for 4 pancakes is 240 mT, while the minimum one for a single pancake is 160 mT. Then the minimum $n$ for all the measured cases is that for 240 mT at $90^{\rm o}$ and the maximum one is that for 160 mT at $0^{\rm o}$. Using the extrapolations from section \ref{s.IcBath}, we find that $n$ is between 18.3 and 28.1. With these $n$ values, the critical current for 0.1 $\mu$V/cm is 0.88 and 0.92 times $I_c$ at 1 $\mu$V/cm, respectively. The difference between both curves is $4\%$, and thence the error in the average value by taking an approximated $n$ is only 2\%. The gray area in figure \ref{f.Icnp} represents the region between the two limits. The mid value between these limits agrees well with the experiments. The maximum and minimum differences are 6.4 and 0.9 \% for 3 and 1 pancakes, respectively.

Simulations are useful to predict situations difficult to measure, such as the effect of adding many pancakes to the stack (figure \ref{f.Icnp}). The critical current decreases with increasing the number of pancakes until it saturates. This is because (i) the top and bottom coils limit the critical current and (ii) the magnetic field there saturates with increasing the coil height.


\subsection{AC loss calculations}
\label{s.losscalc}

In the following, we first validate the AC loss simulations by comparing to the experiments and afterwards we compute situations which are difficult to measure. These are the AC loss in individual pancakes of a stack, the effect of adding more pancakes and the AC loss reduction by a 2-strand Roebel cable.

\subsubsection{Comparison with experiments}

\begin{figure}[tbp]
\centering
\includegraphics[width=8 cm]{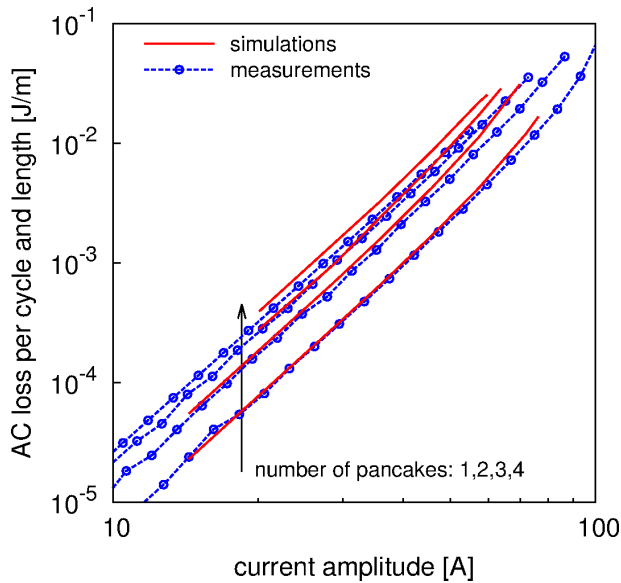}
\caption{\small Simulations of the AC loss per cycle and unit tape length (solid red lines) together with the experiments (dash blue lines with symbols). The number of pancakes are 1,2,3,4 in the arrow direction.}
\label{f.Qcomp}
\end{figure}

Simulations agree well with experiments in most cases (figure \ref{f.Qcomp}). However, the simulations overestimate AC loss, increasing the deviation with approaching to the critical current. This is due to the sharp $E(J)$ relation of the critical-state model \cite{roebelcomp,chen05APL}. The best agreement is for one pancake. The maximum deviation for current amplitudes below 0.7 times the critical current of the coils is 6\% for a single pancake, 15\% for stacks of 2 and 3 pancakes and 25\% for 4 pancakes. The deviation for the largest number of pancakes could be due to the imperfect formula for $J_c(B,\theta)$, see figure \ref{f.IcBath}. Our formula fits experimental data up to 200 mT but the local magnetic field in the 4-pancake coil is up to 250 mT. Then, there could be some error in the extrapolation. For one pancake at currents below 0.5 times the critical current, the discrepancy with the measurements is around 3\%. This agreement is compatible with that between the measured and simulated critical current of the tape, figure \ref{f.IcBath}. From that comparison, we infere that the average deviation of the internal $J_c$ is below 4\%. A discrepancy of only 3\% in the AC loss can be explained by the inevitable experimental error in the measurements of the AC loss and the tape critical current. Simulations for current amplitudes above the critical current of the coil would require a model with a smooth $E(J)$ relation.

\subsubsection{AC loss in each pancake of a stack}
\label{s.losspanc}

\begin{figure}[tbp]
\centering
\includegraphics[width=8 cm]{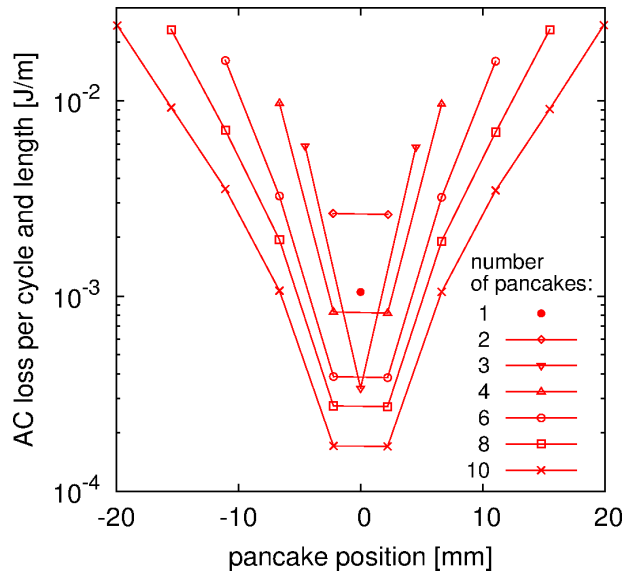}
\caption{\small Simulations show that the AC loss per cycle and unit tape length in each pancake (or the AC loss per pancake) increases with approaching to the end of the stack. The current amplitude is 46.0 A for all cases.}
\label{f.Qpan}
\end{figure}

In contrast to global electrical measurements, simulations can determine the local AC loss. This is because we calculate the AC loss by the integration of ${\bf J}\cdot{\bf E}$, which is the local power dissipation.

The AC loss (or the AC loss per tape length) in the central pancake is the smallest (figure \ref{f.Qpan}). Actually, the AC loss in the central pancake for stacks of 3 pancakes or more is smaller than when that pancake is alone. This is because for stacks of pancakes the magnetic field at the centre is roughly parallel to the tapes, while for a single pancake there is an important perpendicular component (figure \ref{f.rA}). The larger the number of pancakes, the lower the AC loss in the central ones. This is because for an infinitely long coil, the field is parallel to the tapes. However, there will always be a small perpendicular component due to the gap between pancakes.

For 3 pancakes, the AC loss at the central pancake is smaller than the AC loss in the central pancakes for 4 pancakes, figure \ref{f.Qpan}. This is because for 3 pancakes the maximum perpendicular field in the central one is smaller than for 4 pancakes. In addition, the vertical symmetry of the coil causes that the critical regions for 3 pancakes penetrate from the top and bottom of the pancake while for 4 pancakes they penetrate practically only from one side (see figure \ref{f.Jcen}). As a consequence, the distance of the penetration depth is smaller for 3 pancakes. For 4 pancakes, both the penetration depth and the maximum magnetic field are larger, so the AC loss in the central tape is larger. In more detail, the power AC loss is the integration of $JE$ in the superconducting section. At the undercritical region, $E=0$, and thence from the Faraday law it follows that $E\propto -\partial\Phi/\partial t$, where $\Phi$ is the flux between the undercritical region and the point where it is evaluated and $t$ is the time. Since at the central turns for 4 pancakes, both the maximum magnetic field and the penetration depth are larger than for 3 pancakes, the flux is larger and so is the AC loss. This reasoning can be extended to any given odd number of pancakes and the following even number of pancakes.

The largest AC loss per tape length is at the pancakes on the ends (figure \ref{f.Qpan}). The reason is that the perpendicular magnetic field is the largest there (figure \ref{f.rA}). The AC loss per tape length in those pancakes saturates for coils with a large enough number of pancakes.

These results have implications in cryogenics. Most of the heat is generated at the ends of the coil, where it is the easiest to extract, either by a cryogenic fluid or by conduction cooling. Heat extraction at the inner pancakes is usually more difficult. However, the heat generated there will be the smallest. Another advantage of the fact that most of the AC loss is generated at the top and bottom of the coils is that it can be reduced by means of magnetic diverters \cite{pancakeFM,shevchenko98PhC,almosawi01IES,furuse04IES}.

\subsubsection{Dependence on the number of pancakes}
\label{s.lossnumb}

\begin{figure}[tbp]
\centering
\includegraphics[width=8 cm]{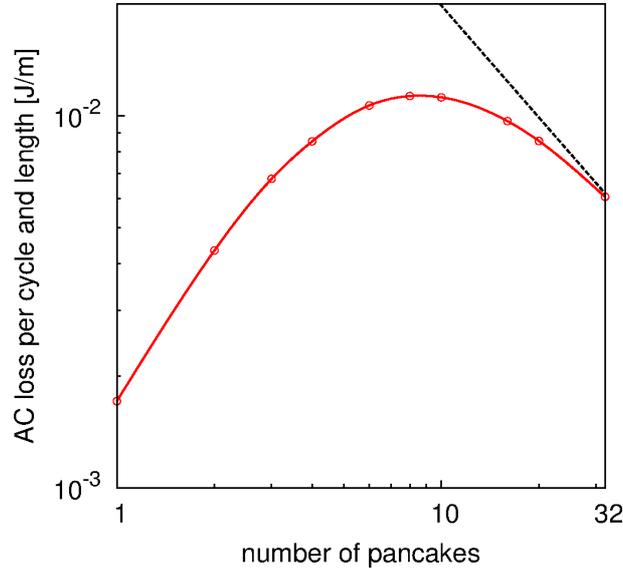}
\caption{\small The computed AC loss per cycle and tape length presents a peak with increasing the number of pancakes (solid red line with symbols), given a certain current amplitude (46.0 A in this case). The dash black line represents a function inversely proportional to the number of pancakes.}
\label{f.Qmany}
\end{figure}

At a fixed current, the AC loss per unit tape length increases with increasing the number of pancakes until it reaches a maximum, decreasing for large number of pancakes (figure \ref{f.Qmany}). A decrease of a power -1 would mean that the total AC loss of the coil is saturated. The AC loss per unit tape length (or per pancake) for the highest number of pancakes still does not decrease as a power -1. This is because the height-to-diameter ratio of the coil is not very large (around 2) and the perpendicular component of the magnetic field is not negligible in most of the coil. For very large number of pancakes, the AC loss per unit tape length will saturate at a value much smaller than the one for a single pancake. This is because the magnetic field far away from the ends will be practically parallel to the tapes, although there will always be a perpendicular component due to the gaps between the pancakes.

The AC loss per unit tape length in a double pancake is around 3 times larger than a single one (figure \ref{f.Qmany}). This is smaller than the expected factor 4 for thick pancakes with large radius and constant $J_c$. The latter factor is obtained as follows. For thick enough pancakes, the AC loss approaches to that of a slab with the same width, $a$, thickness, $b$, and engineering current density $J_{c,e}$ \cite{pancaketheo}. The loss per length of the slab is $Q=\mu_0b^3aJ_{c,e}^2(I_m/I_c)^3/6$, where $I_m$ is the current amplitude \cite{norris70JPD}. Then, for a coil with internal radius much larger than its width and high number of turns, the AC loss per tape length is
\begin{equation}
Q=\frac{\mu_0w^3d^2J_c^2(I_m/I_c)^3}{6(d+h)},
\end{equation}
where $w$, $d$ and the width and thickness of the superconducting layer, and $h$ is the separation between superconducting layers of neighbouring turns. As a consequence, doubling the width of the coil results in 8 times larger loss. A double pancake coil behaves as a single one with double width but it has double the number turns. Then, the loss per tape length in a double pancake is 4 times larger than a single one. For our case, placing two coils together increases less the loss per length because the coil thickness-to-height aspect ratio decreases, so the local magnetic field increases in a lower proportion than placing together infinitely thick pancakes.

\subsubsection{Coils made of Roebel cable}
\label{s.lossRoeb}

\begin{figure}[tbp]
\centering
\includegraphics[width=8 cm]{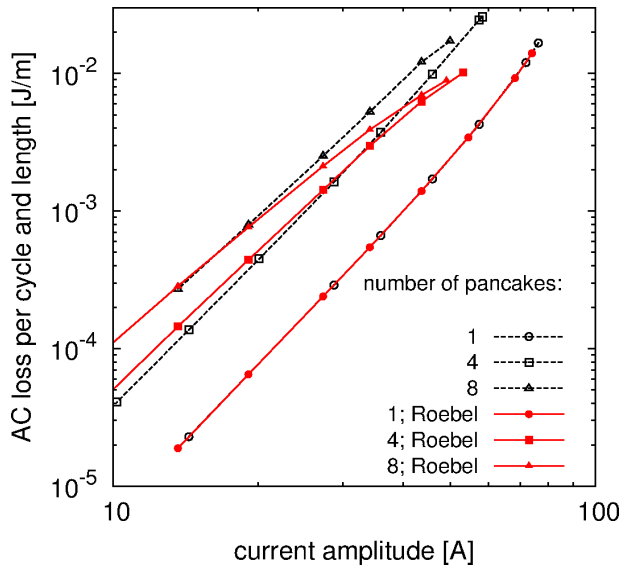}
\caption{\small Simulations reveal that replacing the tape by a 2-strand Roebel cable of the same width reduces the AC loss in stacks of pancakes but not in a single pancake (section \ref{s.lossRoeb}). The plot shows the AC loss per cycle and tape/cable length. The Roebel cable decreases slightly the critical current.}
\label{f.QRoeb}
\end{figure}

Roebel cables are promising in order to reduce the AC loss due applied magnetic fields \cite{J:2006:Goldacker06,J:2008:Long08a,J:2009:LeeJK09a}. These cables are made of transposed strands. Our experimental coils are made of 4 mm tapes. In this section, we study the case that the Roebel cable is made from cutting 4 mm wide tapes, so the cable is made of two strands and the total width is the same as the original tape. Then, the critical current of the cable is slightly smaller than the original tape because of the loss of superconducting material. We take the approximation that the transposition length is much larger than the cable width. Then, the crossover segments are much smaller than the longitudinal segments in the strands, and thus the AC loss per cable length is approximately that of the parallel segments. As a consequence, we can use our mathematically 2D model \cite{roebelmeassim,roebelcomp}. Since in a Roebel cable with electrically isolated filaments the net current is the same in all strands, we can simulate a coil made of a 2-strand Roebel cable as a coil with double number of pancakes. In order to simplify the calculations, we take the same separation between pancakes as the gap between strands in the Roebel cable. For the simulations, we use the same dimensions as the measured pancake coils (table \ref{t.dimcoils}) and a separation between pancakes or gap between Roebel strands of 200$\mu$m. We take the anisotropic field-dependent $J_c$ from equations (\ref{Jcall})-(\ref{fpi}).

The simulations show that a Roebel cable does not reduce the AC loss in a single pancake coil (figure \ref{f.QRoeb}). This is because the pancake with a Roebel cable behaves like a double pancake of the same width as the original coil. Since double pancakes roughly behave as a single one with double width (figure \ref{f.rA}), the pancake with Roebel cable behaves as a single one. The critical current with Roebel cable is slightly lower, which should increase the AC loss. However, this loss increase is roughly compensated by the reduction due to the gap in between strands \cite{tranarr}. 

The Roebel cable reduces the AC loss in stacks of pancakes for high currents (figure \ref{f.QRoeb}). This is because the penetration depth of the critical current is reduced. Close to the critical current, the loss decreases down to around 60\% of the original one for both the 4-pancake and 8-pancake stacks. This is close to 50\%, the expected maximum reduction for a Roebel cable in perpendicular applied field \cite{roebelmeassim}. The Roebel cable slightly increases the AC loss at low amplitudes because of the reduction of the critical current.



\section{Summary and conclusion}
\label{s.concl}

This article provides a complete study of the electromagnetic properties of $Re$BCO pancake coils and stacks of them by means of experiments and simulations. In particular, we have studied the AC loss, the critical current and the field and current distribution. First, we have experimentally characterised the coils by electrical means. Afterwards, we have numerically simulated them. In order to do this, we have obtained anisotropic field dependence of the {\em internal} critical current density, $J_c(B,\theta)$, from measurements of the critical current, $I_c$, of the tape. Actually, we have obtained an analytical fit for $J_c(B,\theta)$ by comparing the measured and calculated $I_c$ so the self-field effects in the measurements are taken into account. Both the calculated critical current and AC loss of the coils agree with the measurements, specially the AC loss for a single pancake coil. We have also numerically studied the main properties of stacks of many pancake coils, up to 32. Finally, we have predicted the AC loss reduction in the coils by replacing the tape by a 2-strand Roebel cable of the same width.

The measurements have revealed the following features. First, the hysteresis AC loss per unit tape length increases with the number of pancakes in the stack from 1 to 4. The contribution from the resistive contacts between pancakes is negligible. The critical current of all the pancakes at self-field is roughly the same. However, one of the pancakes (P3) exhibits around 10\% lower critical current when it is under the influence of the field created by the other pancakes. This evidences non-homogeneity in the field dependence of $J_c$. In addition, $J_c$ of the tape does not follow a 180$^{\rm o}$ symmetry and has two clear contributions: isotropic pinning centres and columnar ones in the $c$ direction.

The simulations of the magnetic field and current distribution have shown that a double pancake roughly behaves as a single one with double width. For stacks of 3 or more pancakes, the situation is very different. These stacks of pancakes present magnetisation currents; that is, currents with opposite sign to that of the net current for the peak of the AC cycle. With increasing the number of pancakes, both the critical current and AC loss saturate for a large enough number of pancakes. Indeed, the AC loss at the pancakes far away from the coil ends is negligible, although it never vanishes because of the presence of the gaps between pancakes.

In conclusion, the hysteresis loss in stacks of $Re$BCO pancake coils is very large. Our simulations are useful to predict the AC loss in pancake coils and stacks of them. The predictions for the local dissipation are valuable for the cryogenic design. Effective heat removal is specially important close to the coil ends, where most of the heat is generated. A 2-strand Roebel cable is enough to reduce the AC loss in stacks of pancakes to 50-60\% of the original one for high enough current amplitudes. However, this Roebel cable does not reduce the AC loss for single pancake coils. As future work, we propose to measure and simulate larger coils, coils made of multi-strand Roebel cable and optimisation computations for minimum AC loss or maximum generated field.


\section*{Acknowledgement}

This work was supported by the Structural Funds of the European Union through the Agency for the Structural Funds of the European Union from the Ministry of Education, Science, Research and Sport of the Slovak Republic under the contract number 26240220028.


\bibliographystyle{unsrt}	
\bibliography{pancake_stack.bib}



\end{document}